\documentclass[sigconf]{acmart}

\usepackage{amsmath,amssymb,amsfonts}
\usepackage[ruled,vlined,linesnumbered]{algorithm2e} 
\usepackage{graphicx}
\usepackage{subcaption}
\usepackage{textcomp}
\usepackage[table]{xcolor}
\usepackage{multirow}
\usepackage{import}
\usepackage{enumitem}
\usepackage{xspace}
\usepackage{cleveref}
\usepackage{lipsum}
\usepackage{booktabs}
\usepackage{tabularx}
\usepackage{balance}

\copyrightyear{2026}
\acmYear{2026}
\setcopyright{cc}
\setcctype{by}
\acmConference[FSE Companion '26]{34th ACM Joint European Software Engineering Conference and Symposium on the Foundations of Software Engineering}{July 05--09, 2026}{Montreal, QC, Canada}
\acmBooktitle{34th ACM Joint European Software Engineering Conference and Symposium on the Foundations of Software Engineering (FSE Companion '26), July 05--09, 2026, Montreal, QC, Canada}
\acmDOI{10.1145/3803437.3805251}
\acmISBN{/2026/07}

\newboolean{showcomments}
\setboolean{showcomments}{true} 
\ifthenelse{\boolean{showcomments}}{
  \newcommand{\nbc}[3]{
    {\colorbox{#3}{\bfseries\sffamily\scriptsize\textcolor{white}{#1}}}%
    {\textcolor{#3}{\sf\small$\blacktriangleright$\textit{#2}$\blacktriangleleft$}}}
}{
  \newcommand{\nbc}[3]{}
}

\definecolor{mygreen}{rgb}{0.15, 0.65, 0.25}
\definecolor{myred}{rgb}{0.72, 0.13, 0.13}

\def\BibTeX{{\rm B\kern-.05em{\sc i\kern-.025em b}\kern-.08em
    T\kern-.1667em\lower.7ex\hbox{E}\kern-.125emX}}
    
\AtBeginDocument{%
  \providecommand\BibTeX{{%
    Bib\TeX}}}

\begin{document}

\author{Shuai Wang}
\affiliation{%
  \institution{Chalmers University of Technology}
  \city{Gothenburg}
  \country{Sweden}}
\email{shuaiwa@chalmers.se}

\author{Yinan Yu}
\affiliation{%
  \institution{Chalmers University of Technology}
  \city{Gothenburg}
  \country{Sweden}}
\email{yinan@chalmers.se}

\author{Earl Barr}
\affiliation{%
  \institution{University College London}
  \city{London}
  \country{United Kingdom}}
\email{e.barr@ucl.ac.uk}

\author{Dhasarathy Parthasarathy}
\affiliation{%
  \institution{Volvo Group}
  \city{Gothenburg}
  \country{Sweden}}
\email{dhasarathy.parthasarathy@volvo.com}

\title[LLM-Powered Workflow Optimization for Multidisciplinary Software Development]{LLM-Powered Workflow Optimization for Multidisciplinary Software Development: An Automotive Industry Case Study}

\begin{abstract}
Multidisciplinary Software Development (MSD) requires domain experts and developers to collaborate across incompatible formalisms and separate artifact sets. Today, even with AI coding assistants like GitHub Copilot, this process remains inefficient; individual coding tasks are semi-automated, but the workflow connecting domain knowledge to implementation is not. Developers and experts still lack a shared view, resulting in repeated coordination, clarification rounds, and error-prone handoffs. We address this gap through a graph-based workflow optimization approach that progressively replaces manual coordination with LLM-powered services, enabling incremental adoption without disrupting established practices. 
We evaluate our approach on \texttt{spapi}, a production in-vehicle API system at Volvo Group involving 192 endpoints, 420 properties, and 776 CAN signals across six functional domains. The automated workflow achieves 93.7\% F1 score while reducing per-API development time from approximately 5 hours to under 7 minutes, saving an estimated 979 engineering hours. In production, the system received high satisfaction from both domain experts and developers, with all participants reporting full satisfaction with communication efficiency.
\end{abstract}


\maketitle

\section{Introduction}

In many industrial software projects, domain experts and software developers must collaborate across disciplinary boundaries, coordinating through heterogeneous artifacts that each party produces, maintains, and evolves independently. Experts contribute specifications, signal definitions, and design documents grounded in their respective domains, while developers must reconcile these artifacts and translate them into a unified, executable codebase. This coordination pattern is straightforward when artifacts are few and stable, but becomes a persistent bottleneck as systems grow in scope and the number of contributing disciplines increases. The automotive industry, with its deep regulatory requirements, safety-critical constraints, and multidisciplinary teams, offers an instructive example of how document-to-code translation can dominate engineering effort \cite{sam2025generating}.

\begin{figure}[t!]
    \centering
    \includegraphics[width=1.0\linewidth]{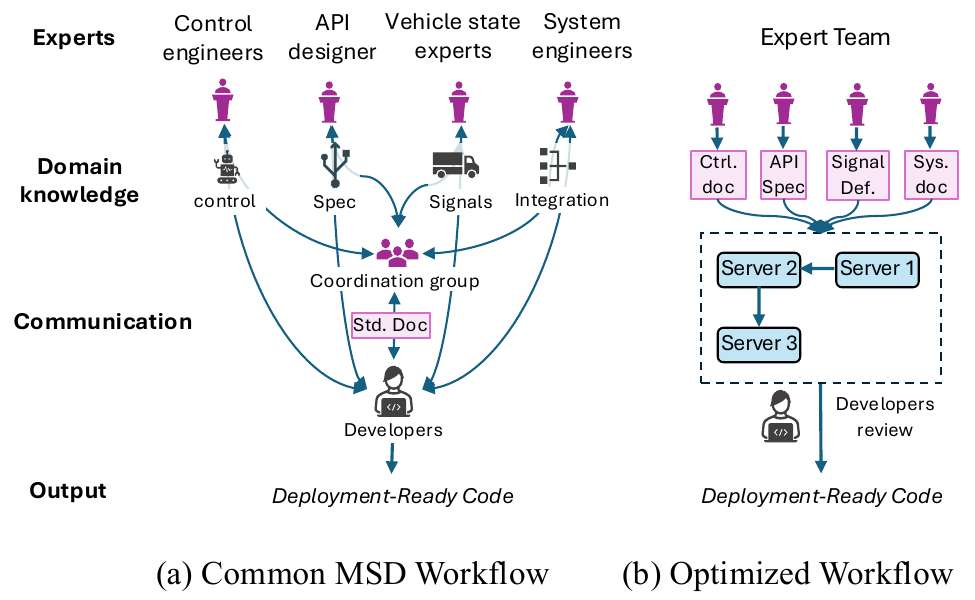}
    \caption{Comparison of workflows in the Multidisciplinary Software Development (MSD) process: (a) a typical MSD workflow and (b) optimized workflow through automated translation via our graph-based approach.}
    \label{fig:overview}
\end{figure}

This pattern is characteristic of Multidisciplinary Software Development (MSD), which arises whenever software systems must encode specialized knowledge from non-software domains. In MSD settings, domain experts (engineers, scientists, regulators, or business analysts) produce specifications using discipline-specific formalisms that software developers must interpret, align, and implement as coherent system behavior~\cite{li2017cross}. The collaboration is inherently two-way: developers translate domain requirements into code, while domain experts must understand implementation constraints to refine their specifications~\cite{KBE1,KBE2}. This continuous dialogue creates value by embedding deep expertise into software, but it also introduces persistent operational challenges that scale poorly with system complexity.

This paper presents a practical case study in automotive software development at Volvo Group. We use \texttt{spapi} as a running example: an in-vehicle web server that exposes vehicle state and control-related properties through RESTful APIs to driver-facing mobile applications as well as backend services. Developing \texttt{spapi} requires a multidisciplinary team, including product owners, business translators, UI and app designers, Android engineers, software developers, control engineers working in MATLAB/Simulink, system engineers, and system architects. These roles rely on distinct technical vocabularies, toolchains, and artifact formats, and changes in one domain often propagate to others in ways that are difficult to anticipate or track. Using this case study, we highlight recurring operational problems in translating domain artifacts into code and show how LLM-based automation can substantially reduce manual overhead while maintaining delivery quality.

\Cref{fig:overview} illustrates a common pattern we observe in MSD: requirements are authored across heterogeneous documents by different stakeholders, then manually translated by developers into implemented endpoints, followed by repeated feedback cycles where feasibility and constraints are communicated back to domain experts for refinement. In \texttt{spapi}, this translation spans a patchwork of documents and a correspondingly tedious patchwork of implementation tasks, which slows delivery and increases the risk of defects. Our goal, shown in \Cref{fig:overview}(b), is to automate the translation from domain artifacts to \texttt{spapi} endpoints. Achieving this requires transforming the MSD workflow itself: we reorganize the handoffs and dependencies so that LLM-powered services can replace repeated manual coordination. We design this workflow transformation using a graph-based representation of the artifacts and their dependencies.
We make four contributions relevant to practitioners facing similar MSD challenges:

\begin{itemize}
\item We showcase the operational reality of artifact-driven multidisciplinary development at scale by characterizing the translation bottleneck in \texttt{spapi}, an in-vehicle API system at Volvo Group. The failure modes we identify generalize across industries where software must encode specialized domain knowledge, offering practitioners a vocabulary for diagnosing similar inefficiencies in their own workflows.

\item We show that automating translation requires transforming the MSD workflow itself, not just accelerating individual coding tasks. We model the workflow as a graph of artifacts, dependencies, and handoffs, and use this representation to systematically restructure the process so that LLM-powered services can incrementally replace manual coordination while keeping domain experts in the loop and minimizing disruption to established practices.

\item We provide quantitative evidence from a deployed system that generates 192 real-world API endpoints across six industrial domains. Compared to semi-automated implementations (human developers assisted by GitHub Copilot), the automated workflow achieves a 93.7\% F1 score and reduces per-API development time from approximately 5 hours to under 7 minutes, saving an estimated 979 engineering hours across the endpoint portfolio.

\item We report deployment experience and stakeholder feedback from production use at a major automotive manufacturer. Domain experts and developers using the system daily reported high satisfaction (averaging 4.80 and 4.67, respectively, on a 5-point scale), indicating that the approach delivers practical value beyond offline metrics.
\end{itemize}

The rest of this paper is organized as follows. Section~\ref{sec:spapi} describes the \texttt{spapi} system and its workflow challenges. Section~\ref{sec:methods} presents our optimization approach. Section~\ref{sec:evaluation} reports results and stakeholder feedback. Section~\ref{sec:discussion} discusses practical considerations, Section~\ref{sec:related} reviews related work, and Section~\ref{sec:conclusion} concludes.
\section{The \texttt{spapi} System}
\label{sec:spapi}
\texttt{spapi} is an in-vehicle web server deployed at Volvo Group that provides an authenticated API layer mapping vehicle signals and control state to well-typed RESTful endpoints. Driver-facing mobile applications and backend fleet services consume these endpoints to display vehicle status, adjust comfort settings, and monitor operational parameters across the vehicle's functional domains.


Developing and maintaining \texttt{spapi} has required 15 to 20 full-time engineers (FTEs) to deliver more than 100 APIs across six functional domains: driver productivity, connected systems, energy management, vehicle systems, visibility, and dynamics. Three types of artifacts drive this workflow. OpenAPI specifications define endpoint behavior, data types, and interface contracts; they capture the essential complexity of the system and serve as the authoritative source for mobile and cloud integration~\cite{DBLP:journals/computer/Brooks87}. Signal definitions describe low-level CAN bus messages that encode vehicle state, authored by control engineers and system architects with knowledge of the physical systems. Mapping documents bind high-level API properties to their underlying signals, specifying how abstractions like ``climate mode'' translate to specific CAN message fields.

The typical workflow proceeds through a sequence. Domain experts produce requirements and signal descriptions based on vehicle capabilities and business needs. Developers manually convert those documents into endpoints, adapters, data models, and tests, consulting the various artifact types to understand how each API property should behave. A dedicated coordination group, supplemented by direct contact with domain experts, resolves ambiguities that arise when specifications are incomplete or inconsistent. Under this workflow, defining a single API could take as long as 10 weeks from initial specification to deployment-ready code.

\paragraph{Observed Problems}
Four problems characterize the workflow's failure modes, each contributing to inefficiency that compounds as the system scales.

\textbf{Fragmented artifacts.} Multiple overlapping documents, including API specifications, signal definitions, detailed descriptions, and coordination notes, describe related information in different formats and at different levels of abstraction. Developers must cross-reference and align these sources to produce correct implementations. When artifacts fall out of sync, as they frequently do during requirements evolution, the reconciliation burden compounds and error opportunities multiply.

\textbf{Manual-translation overload.} Developers spend substantial effort transcribing information from specifications and signal definitions into code, drawing time away from higher-value work such as design, testing, and verification. Even with coding agents like GitHub Copilot, they must still locate relevant artifacts, interpret domain-specific formalisms, reconcile inconsistencies, and consult domain experts to resolve ambiguities. This burden scales with API and signal count. With over 400 unique properties and 776 associated CAN signals in the \texttt{spapi} system, translation remains a persistent bottleneck that coding assistants alone cannot remove; the workflow itself must be automated.

\textbf{Ambiguity-driven churn.} Underspecified domain documents trigger repeated clarification between developers and experts, consuming time and sometimes invalidating earlier implementation decisions. Signal definitions in \texttt{spapi} were often so concise as to be ambiguous, requiring consultation with vehicle state experts, system engineers, and sometimes control engineers, whose discipline has deep technical foundations spanning decades~\cite{10.5555/583119}. A single ambiguous signal could delay implementation by days.

\textbf{Coordination bottlenecks.} The coordination group created to streamline information flow in \texttt{spapi} development proved insufficient for managing hundreds of API properties and vehicle signals. Developers began bypassing it to contact domain experts directly, creating ad hoc communication patterns that were hard to track and prone to inconsistency. Once such workarounds start delivering value, organizational inertia makes them difficult to replace, even when their inefficiency is widely recognized.

\paragraph{Industry Implications}

These problems are not unique to automotive systems. Similar bottlenecks arise wherever software must encode specialized domain knowledge and development requires ongoing coordination across disciplinary boundaries.

The consequences compound at scale. Manual transcription shifts engineering time away from architecture, testing, and verification, increasing both development costs and defect risk. Fragmented artifacts and frequent clarification cycles produce inconsistent implementations, with conflicts discovered late in integration when rework costs are highest. As product scope or regulatory requirements grow, staffing and coordination overhead increase disproportionately, yet adding engineers provides only temporary relief without addressing underlying inefficiencies. Perhaps most critically, domain understanding becomes concentrated in role-specific artifacts and individual expertise; when personnel change or specifications evolve, the implicit knowledge embedded in translation decisions can be lost, forcing costly rediscovery.

These failure modes motivate automation strategies that reduce manual handoffs through tighter artifact-code integration and targeted tooling. The following sections describe our approach and its application to \texttt{spapi}.

\paragraph{Our Approach}

We address these challenges through an iterative workflow optimization approach that models the MSD process as a graph and progressively automates manual translation nodes. The workflow graph represents participants (domain experts, developers, and automated services) as nodes, with artifact exchanges as edges. This representation makes explicit the communication structure that drives development effort and highlights where manual translation creates bottlenecks.

We apply a series of graph transformations that replace manual nodes with LLM-powered services, reducing complexity and communication overhead while preserving the information flows that domain experts rely upon. Each transformation targets a specific translation task: generating signal access code from definitions, aligning API properties with corresponding signals, and assembling complete endpoints from validated components. Rather than replacing the entire workflow at once, we incrementally automate individual tasks, validating each step with domain experts before proceeding to the next.

This graph-based perspective enables systematic identification of automation opportunities and provides a framework for measuring improvement across iterations. The result is a production pipeline comprising three coordinated services that reduce per-API development time from approximately 5 hours to under 7 minutes while achieving quality comparable to manual implementation.
\section{Methods and Implementation}
\label{sec:methods}

This section describes our approach to automating artifact-driven multidisciplinary workflows. We model the development process as a graph and apply iterative transformations that replace manual translation tasks with LLM-powered services.

\subsection{Workflow Graph Representation}
To formalize the complex interactions in software development, we represent the workflow as a directed dependency graph $\mathcal{G} = (\mathcal{V}, \mathcal{R})$. We define nodes $\mathcal{V}$ as concrete artifacts (e.g., specification documents, signal definitions, or constraint descriptions) and edges $\mathcal{R}$ as information dependencies.

Formally, let $\mathcal{V} = \{ d_1, d_2, \dots, d_n \}$ denote the set of documents generated during the workflow. A relation $(d_i, r_{ij}, d_j) \in \mathcal{R}$ exists if the production or validation of document $d_j$ depends on the information contained in $d_i$. The workflow $\mathcal{G}$ is thus represented as a set of relational triples:
\begin{equation}
\mathcal{G} = \{ (d_i, r_{ij}, d_j) \mid d_i, d_j \in \mathcal{V}, r_{ij} \in \mathcal{R} \}
\end{equation}
This formulation shifts the problem of workflow automation from managing human coordination to a structured transformation of a document dependency graph into executable code.

\begin{figure*}[h!]
    \centering
    \includegraphics[width=1.0\linewidth]{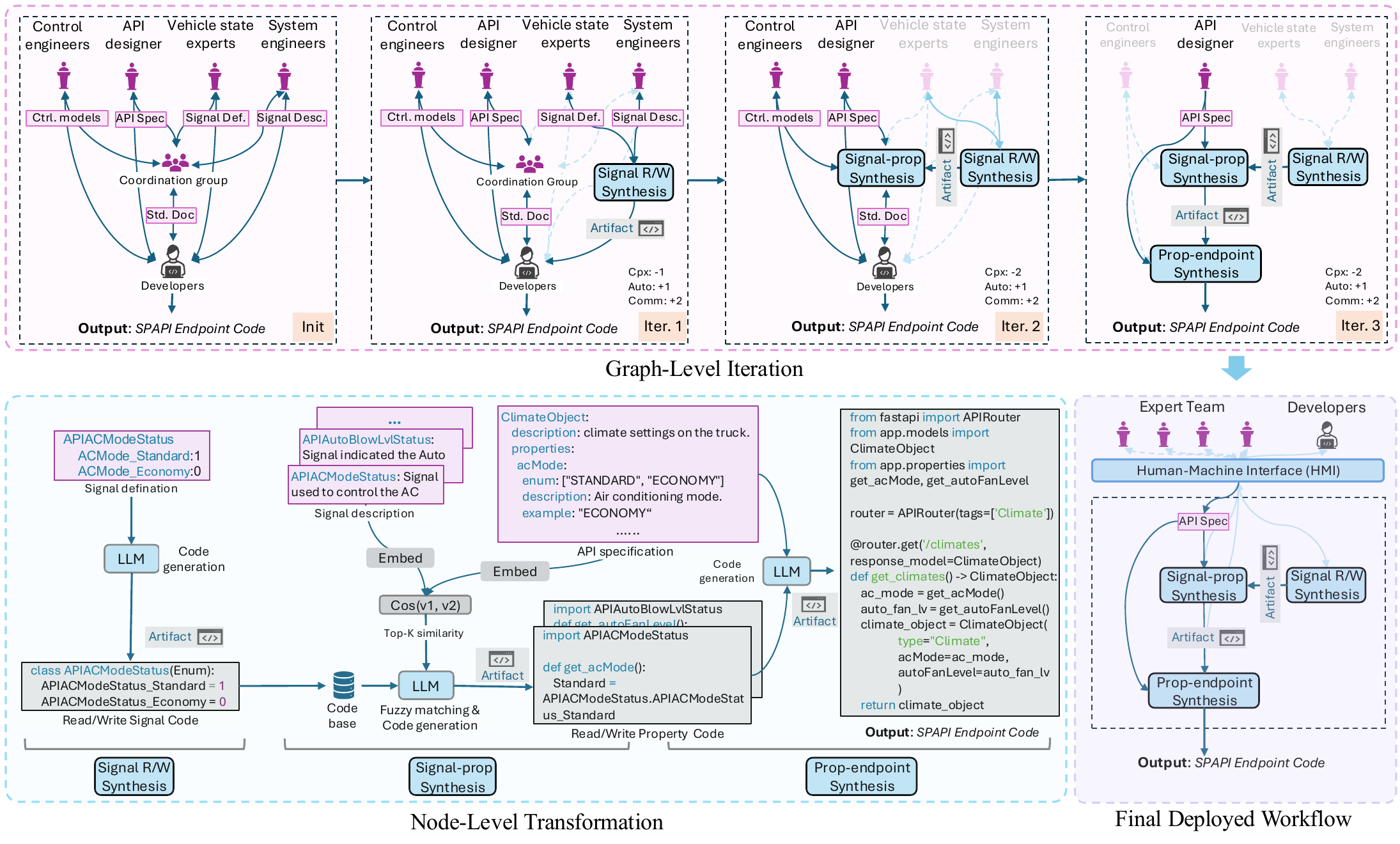}
    \caption{Iterative workflow optimization framework for vehicle API generation. The upper part illustrates graph-level optimization, depicting how the initial manual workflow undergoes three iterative refinements to achieve a highly automated workflow. The lower-right part illustrates the final deployed workflow, which comprises three server nodes. The lower-left part illustrates the detailed structure of each node.}
    \label{fig:polymer-transformation-example}
\end{figure*}

Figure~\ref{fig:polymer-transformation-example} (Init) shows the initial workflow graph for \texttt{spapi}. Four expert roles participate: \emph{API designers} responsible for OpenAPI specifications, \emph{vehicle state experts} who define CAN signal semantics, \emph{system engineers} who integrate signals across subsystems, and \emph{control engineers} who establish vehicle behavior constraints. These experts produce artifacts that developers must translate into code, with a coordination group mediating communication. The graph is densely connected: all four expert roles interact with each other and with developers, resulting in significant coordination overhead.
Without an explicit graph representation, such workflow transformations tend to remain ad-hoc, making coordination bottlenecks difficult to reason about systematically or evaluate across iterations.

This workflow graph $\mathcal{G}$ serves as the starting point for optimization. Our goal is to systematically replace manual translation nodes with automated services, reducing graph complexity and communication overhead while preserving the information flows that domain experts require.

\subsection{Workflow Optimization Framework}
Our framework aims to optimize the workflow by operating at two distinct levels: \emph{node-level transformation}, which automates individual manual tasks, and \emph{graph-level restructuring}, which streamlines the workflow by eliminating redundant coordination dependencies.

\paragraph{Node-level Transformation}
At the node level, we formalize the conversion of manual translation tasks into LLM-driven automated services. For each document node $d_i \in \mathcal{V}$, we denote its associated input materials as $\mathcal{M}_i$ (e.g., domain specifications and signal definitions) and human-provided refinement instructions as $\mathcal{I}_i$. The transformation process is modeled as a function:
\begin{equation}
f_{\text{LLM}} : (\mathcal{M}_i, \mathcal{I}_i) \rightarrow c_i
\end{equation}
where $c_i$ represents a modular, executable code artifact. To ensure reliability, each $c_i$ must pass an automated validation check $\Phi_{\text{test}}$:
\begin{equation}
\Phi_{\text{test}}(c_i, \mathcal{M}_i) \rightarrow \{\text{pass, fail}\}
\end{equation}
where $\Phi_{\text{test}}$ relies on test cases synthesized from $\mathcal{M}_i$. Only artifacts that satisfy $\Phi_{\text{test}} = \text{pass}$ are accepted. Upon successful validation, the manual node $d_i$ is substituted by an automated service node $s_i$, which encapsulates $c_i$ and exposes a programmatic interface. This substitution is denoted as $d_i \Rightarrow s_i$.

\paragraph{Graph-level Restructuring}
Following node-level transformations, the initial graph $\mathcal{G}$ evolves into an intermediate state $\mathcal{G}'$. We then refine the topology by identifying and removing redundant coordination edges. An edge $(v_i, r_{ij}, v_j) \in \mathcal{G}'$ is defined as \emph{redundant} if the information dependency $r_{ij}$ can be resolved through direct programmatic invocation between their corresponding automated services. 

Formally, we define an edge as \texttt{isRedundant} if:
$\exists\, s_i, s_j \;$ such that $d_i \Rightarrow s_i,\ d_j \Rightarrow s_j,\ \text{and } s_j $
can directly consume outputs of $s_i$.

The optimized workflow graph $\mathcal{G}^*$ is reached through an iterative reduction process:
\begin{equation}
\mathcal{G}^{(k+1)} = \{ r \in \mathcal{G}^{(k)} \mid \neg \texttt{isRedundant}(r) \}
\end{equation}
The process converges at a fixed point where $\mathcal{G}^{(k+1)} = \mathcal{G}^{(k)}$. Throughout this reduction, domain experts conduct reviews to ensure that no essential information flow is lost.

\subsection{Measuring Transformation Impact}
To formalize the assessment, we model a transformation \( \tau \) as a mapping from an original workflow \( W \) to a transformed workflow \( W' = \tau(W) \). The overall impact of \( \tau \) is evaluated along the three dimensions defined in Table~\ref{tab:scoring-scheme}.

Let the scoring function for each dimension $ d \in \{\textit{complexity},$ $\textit{automation}, \textit{communication}\} $ be
$s_d(\tau) \in \mathbb{Z}$,
where the range and semantics of \( s_d \) are given directly by the discrete levels in Table~\ref{tab:scoring-scheme}. Each score reflects expert judgment on how the transformation changes the corresponding property when comparing \( W \) and \( W' \).
The transformation impact vector is then defined as
\[
\mathbf{s}(\tau) = \big( s_{\textit{complexity}}(\tau),\; s_{\textit{automation}}(\tau),\; s_{\textit{communication}}(\tau) \big).
\]



\begin{table}[t]
\centering
\caption{Scoring scheme for transformation impact across three dimensions}
\begin{tabularx}{\columnwidth}{l c X}
\toprule
\textbf{Dimension} & \textbf{Score} & \textbf{Description} \\
\midrule
\multirow{3}{*}{Complexity} 
  & -1 & Task or exchange removed or combined \\
  & ~0 & No change \\
  & +1 & Task or exchange added \\
\midrule
\multirow{3}{*}{Automation} 
  & -1 & Task becomes more manual \\
  & ~0 & No change \\
  & +1 & Task becomes more automatic \\
\midrule
\multirow{5}{*}{Communication} 
  & -2 & Less explainable AND less executable \\
  & -1 & Less explainable OR less executable \\
  & ~0 & No change \\
  & +1 & More explainable OR more executable \\
  & +2 & More explainable AND more executable \\
\bottomrule
\end{tabularx}
\label{tab:scoring-scheme}
\end{table}

The first dimension measures graph complexity: whether a transformation adds or removes tasks and communication edges. The second measures automation level: whether work shifts from manual to automatic execution. The third measures communication quality through the lens of \emph{duality}, i.e., the degree to which artifacts are both explainable to humans and executable by machines. Source code with meaningful identifiers and comments exemplifies high duality~\cite{10.1145/3377816.3381720}. Transformations that produce executable code with clear documentation score positively on this dimension.

\subsection{Node-Level Automation}

We implemented three automated services to replace manual translation tasks in the \texttt{spapi} workflow. Each service corresponds to a distinct translation step, and together they form the pipeline shown in the lower portion of Figure~\ref{fig:polymer-transformation-example}.


\subsubsection{Signal R/W Synthesis}

The first service generates Python code for reading and writing CAN signals. It takes signal definitions as input (signal names, data types, value ranges, and enumeration mappings) and produces functions that interface with the vehicle's CAN bus.

\begin{figure}[t]
    \centering
    \includegraphics[width=1.0\linewidth]{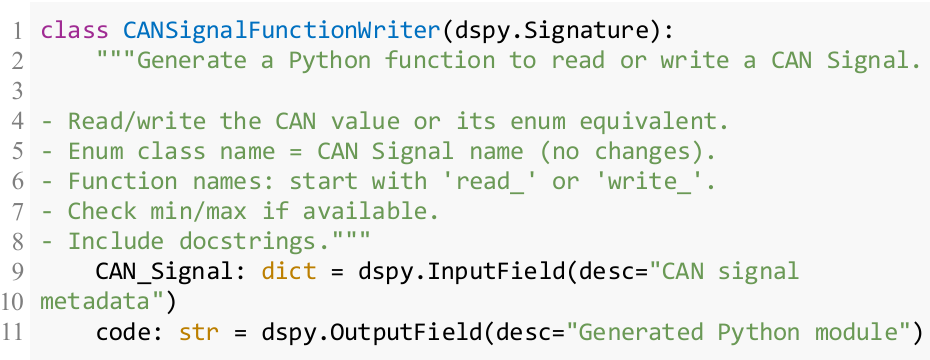}
    \caption{A simplified DSPy signature for Signal R/W Synthesis. The LLM generates Python functions to read or write CAN signals based on signal metadata.}
    \label{fig:dspy-sign}
\end{figure}

We use the DSPy framework~\cite{khattab2023dspy} for structured prompting, as illustrated in Figure~\ref{fig:dspy-sign}. LLMs can be adapted to diverse tasks through prompting~\cite{brown2020languagemodelsfewshotlearners, bubeck2023sparksartificialgeneralintelligence} and excel at processing semi-structured inputs~\cite{jaimovitch2023can, narayan2022can}. DSPy's \texttt{TypedPredictor} ensures that generated code conforms to expected types; if output violates constraints, the system automatically re-prompts the model. 

Recent work on constrained generation~\cite{DBLP:conf/icml/Beurer-Kellner024, park2024grammaraligneddecoding, ugare2024syncodellmgenerationgrammar} enables LLM outputs to satisfy syntactic requirements, facilitating integration with traditional software tooling. For validation, we generate test cases using a separate LLM call and prompt the code-generating LLM to self-debug based on test failures. Once verified, signal R/W code is stored in a vector database indexed by structured signal descriptions, enabling retrieval for downstream synthesis steps. Our test-based validation approach aligns with recent work on LLM-driven test automation in automotive settings~\cite{wang2025automatingcompletesoftwaretest}.

\begin{figure}[t]
    \centering
    \includegraphics[width=1.0\linewidth]{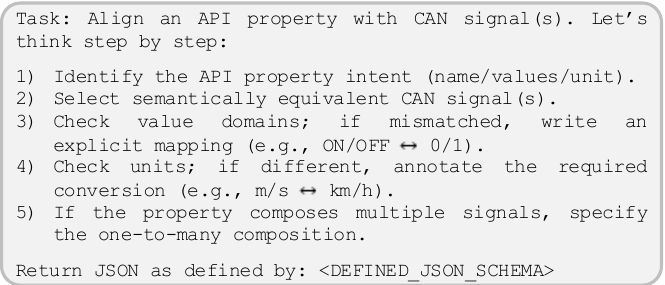}
    \caption{Representative prompt used in the Signal-Property matching task (simplified for clarity).}
    \label{fig:prompt-matching-api}
\end{figure}

\subsubsection{Signal-Property Synthesis}

The second service aligns API properties with their corresponding CAN signal handlers. A key challenge in synthesizing deployable API endpoints is correctly aligning application-level API properties with underlying CAN signals. Given an API property, the system must identify the relevant CAN signals and determine how their values should be interpreted and composed to produce semantically correct behavior.

This is difficult because API and CAN specifications often diverge. Value domains may differ (e.g., \texttt{ON/OFF} vs.\ \texttt{0/1}), units may not match (e.g., \texttt{m/s} on CAN vs.\ \texttt{km/h} in the API), and some properties aggregate multiple CAN signals (e.g., a time property derived from separate minute and second signals). So, naive matching by names or types is unreliable and may introduce silent semantic errors.

To make these assumptions explicit and verifiable, we use a JSON alignment template as an intermediate representation. The LLM must enumerate all contributing CAN signals and specify explicit value mappings, enum correspondences, and unit conversions when needed. As shown in Figure~\ref{fig:prompt-matching-api}, generation is further guided by a chain-of-thought prompt encouraging stepwise reasoning over (i) property intent, (ii) value and unit consistency, and (iii) whether the mapping is direct, transformed, or composed from multiple signals. Together, the template constraint and structured reasoning reduce hallucinated assumptions and improve robustness across heterogeneous signal specifications.

\subsubsection{Property-Endpoint Synthesis}

\begin{figure}[t]
    \centering
    \includegraphics[width=1.0\linewidth]{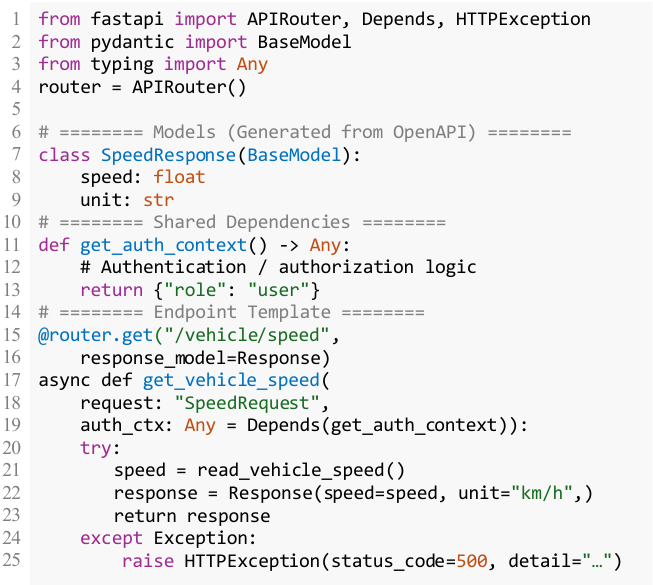}
    \caption{Example of a FastAPI endpoint boilerplate template.}
    \label{fig:boilertemplate}
\end{figure}

\begin{figure}[t]
    \centering
    \includegraphics[width=1.0\linewidth]{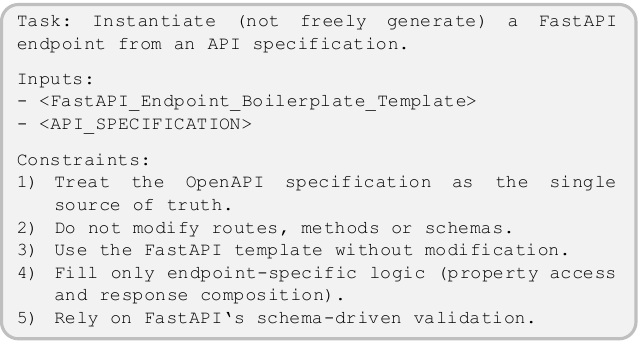}
    \caption{A representative prompt used in the API Endpoint synthesis task (simplified for clarity).}
    \label{fig:prompt-api-endpoint-gen}
\end{figure}

The third service assembles property handlers into deployable API endpoints based on the OpenAPI specification. Given the API contract and generated property accessors, it produces complete FastAPI endpoint implementations, including routing, request validation, response serialization, and error handling.

Synthesizing endpoints directly from specifications is error-prone. Naively generated endpoints may violate the OpenAPI contract by using inconsistent parameter names, omitting or introducing response fields, or mismatching route paths and HTTP methods. Validation logic such as range checks or enum constraints may also be partially implemented or omitted, leading to subtle contract violations that are hard to detect at runtime.

To mitigate these risks, this service adopts a contract-first, template-based generation strategy. A FastAPI endpoint boilerplate template (Figure~\ref{fig:boilertemplate}) fixes endpoint structure, routing, schemas, and cross-cutting concerns such as authentication, logging, and error handling. The LLM is restricted to instantiating endpoint-specific logic within predefined slots, guided by the prompt shown in Figure~\ref{fig:prompt-api-endpoint-gen}. Validation is delegated to FastAPI’s schema-driven mechanisms, ensuring consistent enforcement of constraints specified in the OpenAPI document.

\subsection{Graph-Level Optimization}

Following node-level synthesis, the initial workflow is modeled as a coordination graph $\mathcal{G}^{(0)}$, where nodes represent development activities or automated services, and edges denote information dependencies and human-mediated coordination. Graph-level optimization is an iterative process that restructures $\mathcal{G}^{(k)}$ by introducing executable services to absorb coordination overhead. Formally, each optimization step applies a transformation $\tau_k : \mathcal{G}^{(k)} \rightarrow \mathcal{G}^{(k+1)}$ that replaces manual hand-offs with automated services, thereby eliminating edges resolved programmatically. 

As illustrated in Figure~\ref{fig:polymer-transformation-example}, the transformation of the \texttt{spapi} workflow proceeds through three successive iterations:

\emph{Iteration 1: Signal R/W Synthesis.} By generating executable signal access code directly from definitions, this service eliminates the requirement for developers to consult vehicle state experts, removing the associated coordination edge from $\mathcal{G}^{(0)}$.
    
\emph{Iteration 2: Signal-Property Synthesis.} This stage automates the alignment between API properties and CAN signals. By generating mapping code, the service absorbs the complex interactions previously required between API designers, vehicle state experts, and system engineers, eliminating two additional edges.
    
\emph{Iteration 3: Property-Endpoint Synthesis.} This transformation encodes vehicle-specific constraints (e.g., access permissions and value ranges) directly into OpenAPI annotations. The LLM utilizes these constraints during generation to ensure architectural compliance, removing the need for manual review by control engineers.

The cumulative impact of these transformations is quantified in Table~\ref{tab:iteration-scores}. Each step $\tau_k$ consistently increments the automation level while enhancing communication quality by producing outputs that are both explainable and executable. Collectively, these optimizations reduced the workflow complexity by 5 edges, improved communication quality by 6 units, and increased the overall automation index by 3 relative to the manual baseline.


\begin{table}[t]
\centering
\caption{Cumulative transformation impact across three iterations.}
\begin{tabular}{l c c c}
\toprule
\textbf{Iteration} & \textbf{Comp.} & \textbf{Auto.} & \textbf{Comm.} \\
\midrule
1: Signal R/W  & -1 & +1 & +2 \\
2: Signal-Property & -2 & +1 & +2 \\
3: Property-Endpoint & -2 & +1 & +2 \\
\midrule
\textbf{Net change} & \textbf{-5} & \textbf{+3} & \textbf{+6} \\
\bottomrule
\end{tabular}
\label{tab:iteration-scores}
\vspace{-0.5cm}
\end{table}

\subsection{Deployed System}

After iterative optimization, the final system replaces all manual translation steps with automated services and reaches a stable configuration validated by domain experts. The resulting workflow, shown in the lower-right of Figure~\ref{fig:polymer-transformation-example}, consists of three server nodes in a linear pipeline. Compared with the initial manual workflow, it achieves a net complexity reduction of 5, a communication quality improvement of 6, and an automation increase of 3.

\paragraph{System overview.}
The system takes as input an initial workflow graph $\mathcal{G}^{(0)}$ with dense manual coordination and a set of domain artifacts, including OpenAPI specifications, signal definitions, and constraint documents. It outputs an optimized workflow graph $\mathcal{G}^*$ in which manual translation nodes are replaced by service nodes and redundant coordination edges are removed, yielding a compact executable pipeline.

\paragraph{Architecture and interaction.}
Each node in $\mathcal{G}^*$ is deployed as an independent server with a well-defined API and composed sequentially to mirror the optimized workflow graph. The architecture decouples workflow logic from the underlying language model, enabling substitution of open-source or proprietary LLMs under different cost, latency, and privacy constraints. A unified human--machine interface allows domain experts to inspect API-to-signal mappings, review generated code, and request regeneration with added constraints, preserving human oversight without manual translation.

\paragraph{Execution and adaptability.}
The system supports both fully automated and human-in-the-loop execution. For well-specified tasks, the pipeline runs end-to-end without intervention; ambiguous mappings or specification inconsistencies are flagged for expert review. By replacing input artifacts or domain-specific instructions, the framework can be adapted to new APIs, vehicle platforms, or other translation-heavy workflows.

\section{Evaluation}
\label{sec:evaluation}

\begin{enumerate}[label=RQ\arabic*:]
\item \texttt{[}Overall performance\texttt{]} Does the automated workflow achieve code quality comparable to AI-assisted human development (with Copilot) across different automotive domains?
\item \texttt{[}Reliability\texttt{]} Do the specific techniques employed (boilerplate templates, code composition, and automated debugging) each contribute measurably to system reliability?
\item \texttt{[}Efficiency\texttt{]} Does the automated workflow reduce development time compared to AI-assisted processes while maintaining acceptable quality?
\item \texttt{[}Human factors\texttt{]} Do domain experts and developers report satisfaction with the deployed system sufficient to support continued production use?
\end{enumerate}

\subsection{Experimental Setup}

\begin{table*}[t]
\caption{Performance and satisfaction scores across six automotive domains.}
\centering
\renewcommand{\arraystretch}{1}
\begin{tabular}{p{2.8cm} p{6.8cm} c c c c c c}
\toprule
\multirow{2}{*}{\textbf{Domain}} & \multirow{2}{*}{\textbf{Domain Experts}} & \multirow{2}{*}{\textbf{Num}}  & \multicolumn{3}{c}{\textbf{Performance}} & \multicolumn{2}{c}{\textbf{Satisfaction}} \\
\cmidrule(lr){4-6} \cmidrule(lr){7-8}
& & & \textbf{P} & \textbf{R} & \textbf{F1} & \textbf{Expert} & \textbf{Developer} \\
\midrule
Driver productivity & HMI/UX, business translators, regulators & 34 & 1.000 & 0.957 & 0.978 & 4.93 & 4.86 \\
Connected systems   & Business translators, dealers, sales staff                & 53 & 0.948 & 0.907 & 0.927 & 4.76 & 4.69 \\
Energy              & Control/mechanical/electrical engineers, emissions & 18 & 1.000 & 0.913 & 0.955 & 4.86 & 4.71 \\
Vehicle system      & System architects, embedded software engineers            & 34 & 0.974 & 0.886 & 0.928 & 4.77 & 4.62 \\
Visibility          & Mechanical engineers, designers, HMI/UX        & 14 & 1.000 & 0.925 & 0.961 & 4.88 & 4.75 \\
Dynamics            & Control engineers                                          & 39 & 0.983 & 0.865 & 0.920 & 4.75 & 4.55 \\
\midrule
\textbf{Total}      & -                                                          & 192 & 0.976 & 0.902 & 0.937 & 4.80 & 4.67 \\
\bottomrule
\end{tabular}
\label{tab:domain_performance}
\end{table*}

We conducted experiments on 192 real-world automotive API endpoints collected from six distinct industrial domains at Volvo Group, as summarized in Table~\ref{tab:domain_performance}. These APIs collectively span 420 unique properties and 776 associated CAN signals, representing the full scope of the \texttt{spapi} system.

We use the original production implementations developed by engineers (assisted by GitHub Copilot) as ground truth, referred to as the \emph{baseline}. Precision (P) measures the correctness of generated property-level code, computed as \( P = \frac{|\text{Correct}|}{|\text{Generated}|} \). Recall (R) measures coverage of the original API specification, computed as \( R = \frac{|\text{Correct}|}{|\text{Baseline}|} \). F1 is the harmonic mean of precision and recall, computed as \( F1 = \frac{2PR}{P + R} \).

\vspace{0.3em}

The baseline implementations were reviewed and deployed in production and thus reflect accepted system behavior. In this domain, semantic errors typically appear as enum mismatches, value range violations, or missing signal bindings, all targeted by our validation and automated debugging mechanisms. We therefore adopt a conservative matching strategy, flagging ambiguous cases for human review rather than generating potentially incorrect code.

To address RQ4, we invited all practitioners with direct experience of both workflows to provide structured feedback: four domain experts and two developers. This complete set of qualified evaluators rated the system on role-specific criteria using a 5-point scale. Experts assessed communication effort, functional coverage, and implementation accuracy; developers assessed communication efficiency, debugging effort, and code maintainability.

\paragraph{Baseline workflow.}
Because no existing automated solutions target API development in industrial automotive settings, we compare our system with APIs developed by human engineers assisted by proprietary AI coding agents. In this baseline, engineers were allowed to use GitHub Copilot. We assess both API quality and development time.

\paragraph{LLM usage and time cost.}
The system separates offline processing from deployment-time execution to manage LLM usage efficiently. All compute-intensive LLM operations occur offline and do not affect runtime performance. In our experiments, we use GPT-4o (2024-05-13).

During offline processing, the LLM scans and filters the CAN database and derives read/write equations for relevant signals. In a full run involving 192 APIs and 776 CAN signals, the system made about 15,000 LLM calls over 20.6 hours, averaging 396 seconds and 80 calls per API.

During deployment, the generated read/write equations are reused. The LLM is invoked only for lightweight tasks such as matching API properties to signals and synthesizing the final endpoint. On average, each API requires about 5 LLM calls, taking roughly 10 seconds. This overhead is low enough for practical engineering workflows.

Table~\ref{tab:domain_performance} summarizes code quality results across all six automotive domains. The automated workflow performs strongly throughout: four domains (Driver productivity, Energy, Visibility, and Dynamics) achieve perfect precision of 100\%, meaning all generated property-level code was correct. The other two domains (Connected systems and Vehicle system) still achieve precision above 94\%.

Recall is slightly lower, ranging from 86.5\% to 95.7\% across domains. This reflects our conservative property-to-signal alignment strategy, which applies strict filtering to retain only high-confidence matches. Properties with ambiguous signal mappings are flagged for human review rather than used to generate potentially incorrect code. Even so, average recall exceeds 90\%, indicating broad coverage of the original specifications.

The overall F1 score of 93.7\% shows that the automated workflow produces code quality comparable to baseline development across diverse automotive domains. Performance also remains stable across different expert roles: the system handles APIs requiring input from control engineers (Dynamics, Energy) as effectively as those involving HMI/UX designers or business translators.

Stakeholder satisfaction scores (Section~\ref{sec:rq4}) provide further validation: both experts and developers report high satisfaction and note that the system supports transparent monitoring through well-defined interfaces.

\textbf{Answer to RQ1:} Yes. The optimized workflow achieves 93.7\% F1 with consistent performance across six diverse domains, meeting production quality standards.

\begin{table}[t]
\centering
\caption{Ablation study: impact of individual techniques on code quality and satisfaction.}
\renewcommand{\arraystretch}{1.1} 
\resizebox{\linewidth}{!}{%
\begin{tabular}{lccccc}
\toprule
\multirow{2}{*}{\textbf{Configuration}}& \multicolumn{3}{c}{\textbf{Performance}} & \multicolumn{2}{c}{\textbf{Satisfaction}} \\
\cmidrule(lr){2-4} \cmidrule(lr){5-6}
& \textbf{P} & \textbf{R} & \textbf{F1} & \textbf{Expert} & \textbf{Developer} \\
\midrule
Full automated workflow & 0.976 & 0.902 & 0.937 & 4.80 & 4.67 \\
~~without boilerplate templates & 0.968 & 0.895 & 0.930 & 4.74 & 4.45 \\
~~without code composition & 0.956 & 0.883 & 0.918 & 4.69 & 4.43 \\
~~without automated debugging & 0.911 & 0.841 & 0.875 & 4.33 & 4.15 \\
\bottomrule
\end{tabular}
}
\label{tab:ablation}
\vspace{-0.4cm}
\end{table}

\begin{table}[t]
\caption{Signal code accuracy before and after automated debugging, by signal type.}
\centering
\renewcommand{\arraystretch}{1.1} 
\resizebox{\linewidth}{!}{%
\begin{tabular}{llccccc}
\toprule
\multirow{2}{*}{\textbf{Stage}} & Signal Type & \textbf{Enum} & \textbf{Bool} & \textbf{Numerical} & \textbf{Object} & \textbf{Total} \\
 & Count & 634 & 90 & 33 & 19 & 776 \\
\midrule
\multicolumn{2}{l}{Initial LLM output} & 96.50\% & 96.70\% & 90.90\% & 89.50\% & 93.40\% \\
\multicolumn{2}{l}{After automated debugging} & 100\% & 100\% & 100\% & 100\% & 100\% \\
\bottomrule
\end{tabular}
}
\label{tab:testing_revision}
\vspace{-0.4cm}
\end{table}
\subsection{RQ2: Reliability of Code Generation Techniques}
We conducted an ablation study to quantify the contribution of each technique to overall system reliability. Table~\ref{tab:ablation} shows performance when individual components are removed.

Removing boilerplate templates reduces F1 from 93.7\% to 93.0\% and decreases developer satisfaction from 4.67 to 4.45. Without templates, generated code exhibits inconsistent structure and style, requiring additional manual cleanup. Removing code composition (assembling endpoints from reusable, validated fragments) reduces F1 to 91.8\% and satisfaction scores to 4.69 and 4.43 for experts and developers respectively.

The most significant impact comes from removing automated debugging. Without test-based validation and self-correction, F1 drops to 87.5\%, and satisfaction scores fall to 4.33 and 4.15. This represents a substantial degradation that would likely be unacceptable for production use.

Table~\ref{tab:testing_revision} provides detailed analysis of the debugging component. We categorized the 776 CAN signals into five functional types and measured accuracy before and after the debugging phase. Initial LLM outputs achieve high accuracy for most categories (96.5\% for Enum, 96.7\% for Bool) but lower accuracy for more complex types (90.9\% for Numerical, 89.5\% for Object). Common errors include subtle syntax mismatches, such as using \texttt{LowSupplyPress\textcolor{red}{(}Enum\_Value\textcolor{red}{)}} instead of the correct \texttt{LowSupplyPress\textcolor{green}{[}Enum\_Value\textcolor{green}{]}}. After automated debugging, all signal types reach 100\% accuracy, validating our two-stage strategy of generating an initial draft followed by test-based refinement.

\paragraph{Effect of Embedding Strategies on Property-Signal Matching}

\begin{table}[t]
\caption{Property-to-signal matching performance by embedding strategy.}
\centering
\renewcommand{\arraystretch}{1.1} 
\begin{tabular}{lccc}
\toprule
\textbf{Embedding Input} & \textbf{P} & \textbf{R} & \textbf{F1} \\
\midrule
Raw signal code & 0.763 & 0.277 & 0.406 \\
Original descriptions & 0.861 & 0.451 & 0.592 \\
Rewritten descriptions & 0.980 & 0.925 & 0.952 \\
\bottomrule
\end{tabular}
\label{tab:embedding_results}
\end{table}

We also evaluated the impact of embedding strategy on property-to-signal matching, a critical step in the pipeline. Table~\ref{tab:embedding_results} compares three approaches: embedding raw signal code, embedding original textual descriptions from documentation, and embedding rewritten descriptions enriched with clarified semantics.

All configurations maintain high precision due to strict similarity thresholds, but recall varies substantially. Raw signal code performs poorly (F1 = 0.406) because code lacks semantic context. Original textual descriptions improve performance (F1 = 0.592), but brevity and ambiguity limit accuracy. Rewritten descriptions, where LLMs expand and clarify signal semantics, achieve F1 of 0.952, demonstrating that investment in description quality yields substantial returns.

\subsubsection{Robustness to Specification Errors}

\begin{table}[t]
\caption{Detection performance for common specification errors.}
\centering
\renewcommand{\arraystretch}{1.1} 
\begin{tabular}{lccc}
\toprule
\textbf{Error Type} & \textbf{P} & \textbf{R} & \textbf{F1} \\
\midrule
Out-of-range value  & 0.979 & 1.000 & 0.989 \\
Invalid enum value  & 0.989 & 1.000 & 0.994 \\
\bottomrule
\end{tabular}
\label{tab:error_metrics}
\end{table}

To assess robustness in real-world conditions, we evaluated the system's ability to detect errors in user-provided specifications. We manually injected two common error types into YAML specification files: numerical values exceeding defined ranges and enum values not present in the allowed set.

As shown in Table~\ref{tab:error_metrics}, the system detects these errors with high precision, achieving F1 scores of 0.989 and 0.994 respectively. Beyond ensuring correctness under ideal inputs, the pipeline serves as a validation layer that provides early feedback to help users identify and correct specification errors before they propagate to generated code.

\textbf{Answer to RQ2:} Yes, each technique contributes measurably to system reliability. Automated debugging has the largest impact, improving F1 by 6.2 percentage points and enabling 100\% accuracy on signal code after refinement. Boilerplate templates and code composition each contribute smaller but meaningful improvements. The embedding strategy for property-signal matching also significantly affects performance, with enriched descriptions improving F1 from 0.592 to 0.952.

\subsection{RQ3: Development Time Comparison}

\begin{table}[t]
\caption{Quality and time comparison: automated vs. baseline development.}
\centering
\resizebox{\linewidth}{!}{%
\begin{tabular}{lccccc}
\toprule
\multirow{2}{*}{\textbf{Configuration}} & \multicolumn{3}{c}{\textbf{Performance}} &  \multicolumn{2}{c}{\textbf{Time}}\\
\cmidrule(lr){2-4} \cmidrule(lr){5-6}
 & \textbf{P} & \textbf{R} & \textbf{F1} & Per API & System Total\\
\midrule
Full automated workflow & 0.976 & 0.902 & 0.937 & 396s & 20.6h\\
~~without Signal R/W Synthesis & 0.965  & 0.913  & 0.939  & +1.5h & +288.0h\\
~~without Signal-Property Synthesis & 0.992  & 0.953  & 0.972  & +2.5h & +480.0h\\
~~without Property-Endpoint Synthesis & 0.980  & 0.907  & 0.942  & +1.1h & +211.2h\\
\midrule
Baseline workflow (Copilot) & 0.959 & 0.906 & 0.932 & +5.1h & +979.2h\\
\bottomrule
\end{tabular}
}
\label{tab:synthesis-comparison}
\vspace{-0.5cm}
\end{table}

We compared our automated workflow against baseline development by systematically replacing each automated service with human engineers and measuring both quality and time. Table~\ref{tab:synthesis-comparison} presents the results.

The fully automated workflow completes API generation in 396 seconds per endpoint (approximately 6.6 minutes), while the baseline workflow requires approximately 5.1 additional hours per endpoint, a reduction of over 97\% in development time. Across the full portfolio of 192 APIs, automation saved approximately 979 hours of engineering effort compared to the baseline workflow.

Baseline development achieves higher recall in some configurations, revealing a quality-efficiency tradeoff. When humans perform Signal-Property Synthesis with the help from Github Copilot, F1 increases from 93.7\% to 97.2\% because engineers can resolve ambiguous mappings that the automated system conservatively flags for review. However, this 3.5 percentage point improvement requires 2.5 additional hours per API (480 hours total across the portfolio), a tradeoff that is difficult to justify at scale. 

Each automated service contributes meaningful time savings. Signal R/W Synthesis saves 1.5 hours per API by automating the translation of signal definitions into access code. Signal-Property Synthesis saves 2.5 hours per API -- the largest contribution -- by automating the cognitively demanding task of mapping API properties to signals. Property-Endpoint Synthesis saves 1.1 hours per API by automating endpoint assembly from validated components.

\textbf{Answer to RQ3:} Yes. Per-API development time decreases from 5.1 hours to under 7 minutes (97\% reduction), saving approximately 979 engineering hours across 192 APIs.

\subsection{RQ4: Stakeholder Satisfaction}
\label{sec:rq4}
\begin{table}[t]
\centering
\caption{Detailed satisfaction ratings from domain experts and developers.}
\resizebox{\linewidth}{!}{%
\begin{tabular}{llcc}
\toprule
\textbf{Role} & \textbf{Evaluation Criterion} & \textbf{Score} & \textbf{Average} \\
\midrule
\multirow{3}{*}{Expert (n=4)} 
  & Communication efficiency         & 5.00    & \multirow{3}{*}{4.80} \\
  & Accuracy of implementation       & 4.89 &  \\
  & Coverage of functional requirements & 4.51 & \\
\midrule
\multirow{3}{*}{Developer (n=2)} 
  & Communication efficiency         & 5.00    &  \multirow{3}{*}{4.67}\\
  & Debugging effort                 & 4.37 &  \\
  & Code style and maintainability   & 4.64 & \\
\bottomrule
\end{tabular}
}
\label{tab:human-eval}
\vspace{-0.5cm}
\end{table}

We invited four domain experts and two developers who work with the \texttt{spapi} system to evaluate the deployed workflow. Each participant rated the system on three criteria using a 5-point scale. Table~\ref{tab:human-eval} presents the detailed results.

Both groups report high overall satisfaction, with experts averaging 4.80 and developers averaging 4.67. Most notably, all six participants gave perfect scores of 5.0 for communication efficiency, indicating unanimous recognition that the automated workflow effectively reduces coordination overhead--the primary pain point identified in the baseline process.

Experts rated accuracy of implementation at 4.89, reflecting confidence in the correctness of generated code. The slightly lower score for coverage of functional requirements (4.51) aligns with our quantitative finding that recall is somewhat lower than precision; some edge cases require manual handling. Developers rated code style and maintainability at 4.64, indicating that generated code meets their quality standards, though debugging effort received a slightly lower score of 4.37, suggesting room for improvement in error diagnostics.

The system is deployed in production at Volvo Group and continues to serve as the primary mechanism for API development. Practitioners reported improved transparency and reproducibility compared to the baseline workflow, particularly due to the ability to trace API-to-signal mappings and consistently regenerate code as specifications evolve.

\textbf{Answer to RQ4:} Yes. Despite the small sample size, which represents the complete population of practitioners directly involved in both workflows, all participants reported satisfaction levels supporting continued production use. The unanimous improvement in communication efficiency and sustained production deployment provide converging evidence that the approach effectively addresses the coordination bottleneck motivating this work.

\vspace{-2mm}
\subsection{Threats to Validity}

\textit{Internal validity.} The ground truth may itself contain errors. We mitigate this by using production code that has undergone review and testing. 

\textit{External validity.} Our evaluation focuses on a single API system at one automotive manufacturer. While the 192 endpoints span six functional domains involving diverse expert roles (control engineers, HMI designers, system architects, business translators), generalization to other organizations or industries requires further validation. 

\textit{Construct validity.} Property-level F1 may not capture all aspects of code quality. We supplement quantitative metrics with stakeholder satisfaction ratings to address this limitation.
\section{Discussion}
\label{sec:discussion}

\paragraph{\textbf{On choosing workflows to automate}}
Although Section~\ref{sec:methods} provides technical criteria for workflow automation, selecting suitable workflows ultimately still requires informed judgment. Beyond technical feasibility, successful adoption also depends on organizational readiness, stakeholder alignment, and openness to change, all of which are often difficult to evaluate in advance. In the case of \texttt{spapi}, the workflow was not only well-scoped but also supported by a team willing to experiment and iteratively refine the process, which proved critical to achieving tangible and sustained benefits.
\vspace{-1mm}
\paragraph{\textbf{On the use of LLMs in workflow automation}}
Our results show that LLMs can support multiple stages of workflow automation, including parsing, translation, generation, and evaluation. However, effective use of LLMs still requires experience and engineering judgment, particularly when deciding whether a task should be handled by an LLM alone or by a hybrid solution combining scripts and LLMs. The iterative nature of our approach allows LLM usage to evolve alongside the workflow itself, as teams gradually refine these design choices and improve the balance between automation, control, and reliability over time.
\vspace{-1mm}
\paragraph{\textbf{On multidisciplinary collaboration}}
Multidisciplinary software development often suffers from unclear ownership between developers and domain experts. 
By representing workflows explicitly and enabling LLM-driven translations into domain-accessible forms, our graph-based approach improves transparency and participation. 
This makes workflows more iterable and shareable, helping move teams closer to practical joint ownership.

From a software engineering perspective, our findings suggest that translation-heavy MSD workflows constitute a distinct class of coordination-intensive processes. 
In such workflows, development effort is dominated not by algorithmic complexity, but by repeated artifact translation, handoffs, and clarification cycles across roles. 
Our results indicate that effective automation in these settings lies primarily in restructuring artifact flows and coordination structures, rather than optimizing individual coding tasks. 
We believe this perspective can inform the design of future LLM-assisted tools that operate at the workflow level rather than the function or file level.

\section{Related work}
\label{sec:related}
\paragraph{Improving MSD:}
Although software is increasingly being adopted across diverse fields, research on improving MSD remains relatively limited \cite{wang2025plugging,wang-yu-2025-iquest}.
Previous research includes a survey of non-software engineers~\cite{li2017cross} to understand their perspectives on effective collaboration with software engineers, as well as a study~\cite{feng2025domains} that uses activity theory to interpret sources of friction in MSD.
Other studies have investigated specific types of multidisciplinary collaboration, such as those involving data scientists~\cite{Busquim2024Collaboration, Kim2016DataScientists} and machine learning engineers~\cite{10.1007/978-3-030-19034-7_14, sculley2015hidden,Luo2025HAIEval}, 
emphasizing the nature of these collaborations and the sociotechnical challenges that arise.
Additionally, some research has addressed sector-specific challenges in MSD, such as in healthcare~\cite{6602469} and automotive domains~\cite{doi:10.1142/S0218194019500475, heyn2023automotive}, identifying collaboration issues and their associated costs.

\paragraph{AI in MSD:}
As AI becomes more prevalent in software engineering, recent research has explored how AI can help address collaboration challenges in MSD. 
One interview study \cite{piorkowski2021ai} used the concept of shared mental models to analyze communication gaps between AI developers and domain experts. 
Other studies \cite{subramonyam2025prototyping,zeng2021pan} explored the use of prompting as a means for rapid prototyping among participants with varied workflows. 
Research in Human-Computer Interaction (HCI) has also examined multi-participant interactions with AI systems \cite{Hamza2024Workshop}, 
including the development of guidelines \cite{amershi2019guidelines} and taxonomies \cite{dellermann2021future} for human-AI collaboration. 
Additional works have focused on designing improved interactions \cite{doi:10.1073/pnas.1807184115,Abbasi2025HARESM} to enhance the agency of domain experts.

\paragraph{LLMs for REST APIs:}
In connection with the case study, an expanding body of research leverages LLMs for various aspects of REST API engineering. 
This includes generating API logic from specifications \cite{chauhan2025llm}, creating API documentation from source code \cite{deng2025lrasgen}, producing tests for APIs \cite{wang2025automatingcompletesoftwaretest, kim2024leveraging}, and enabling APIs as tools for LLM applications \cite{bandlamudi2025framework, tao2024harnessing,Alami2025CodeReview}. 

In contrast to prior work that improves isolated tasks, we target the workflow that connects heterogeneous domain artifacts to implemented APIs. We model this MSD workflow as a graph and transform it so LLM-powered services can incrementally replace manual translation and coordination while preserving domain-expert involvement, demonstrated in production on \texttt{spapi}.

\section{Conclusion}
\label{sec:conclusion}

In this paper, we show that LLM-powered workflow automation works in production MSD settings. By modeling workflows as graphs and selectively automating translation steps, we reduce coordination overhead while preserving human oversight. The \texttt{spapi} deployment at Volvo Group validates this approach at scale. The pattern we address is not automotive-specific; it applies wherever software encodes domain knowledge.

\section*{Acknowledgment}
This work was partially funded by the Autonomous Systems and Software Program (WASP), supported by the Knut and Alice Wallenberg Foundation, and the Chalmers Artificial Intelligence Research Centre (CHAIR).
The authors thank Erdem Halil and Shaphan Manipaul Sam Kirubahar for their valuable technical contributions to the research foundations.

\balance
\bibliographystyle{ACM-Reference-Format}
\bibliography{sample-base}

@article{sam2025generating,
  title={Generating APIs through Library Learning using Large Language Models},
  author={Sam Kirubahar, Shaphan Manipaul and Halil, Erdem},
  year={2025}
}

@inproceedings{zeng2021pan,
  title={Pan: Prototype-based adaptive network for robust cross-modal retrieval},
  author={Zeng, Zhixiong and Wang, Shuai and Xu, Nan and Mao, Wenji},
  booktitle={Proceedings of the 44th international ACM SIGIR conference on research and development in information retrieval},
  pages={1125--1134},
  year={2021}
}

@article{wang2025plugging,
  title={Plugging Schema Graph into Multi-Table QA: A Human-Guided Framework for Reducing LLM Reliance},
  author={Wang, Xixi and Costa, Miguel and Kovaceva, Jordanka and Wang, Shuai and Pereira, Francisco C},
  journal={arXiv preprint arXiv:2506.04427},
  year={2025}
}

@inproceedings{wang-yu-2025-iquest,
    title = "i{QUEST}: An Iterative Question-Guided Framework for Knowledge Base Question Answering",
    author = "Wang, Shuai  and
      Yu, Yinan",
    booktitle = "Proceedings of the 63rd Annual Meeting of the Association for Computational Linguistics (Volume 1: Long Papers)",
    month = jul,
    year = "2025",
    url = "https://aclanthology.org/2025.acl-long.760/",
    doi = "10.18653/v1/2025.acl-long.760",
    pages = "15616--15628",
}

@article{Luo2025HAIEval,
  title        = {HAI-Eval: Measuring Human-AI Synergy in Collaborative Coding},
  author       = {Luo, Hanjun and Ni, Chiming and Wen, Jiaheng and Huang, Zhimu and Wang, Yiran and Liao, Bingduo and Chung, Sylvia and Jin, Yingbin and Li, Xinfeng and Xu, Wenyuan and Wang, XiaoFeng and Salam, Hanan},
  journal      = {arXiv preprint},
  year         = {2025},
  url          = {https://arxiv.org/abs/2512.04111},
}

@inproceedings{Hamza2024Workshop,
  title        = {Human-AI Collaboration in Software Engineering: Lessons Learned from a Hands-On Workshop},
  author       = {Hamza, Muhammad and Siemon, Dominik and Akbar, Muhammad Azeem and Rahman, Tahsinur},
  booktitle    = {Proceedings of the ACM/IEEE International Workshop on Software-intensive Business (IWSiB ’24)},
  year         = {2024},
  doi          = {10.1145/3643690.3648236},
}

@article{Abbasi2025HARESM,
  title        = {Towards Human-AI Synergy in Requirements Engineering: A Framework and Preliminary Study},
  author       = {Abbasi, Mateen Ahmed and Ihantola, Petri and Mikkonen, Tommi and M{\"a}kitalo, Niko},
  journal      = {2025 Sixth International Conference on Intelligent Data Science Technologies and Applications (IDSTA)},
  year         = {2025},
  doi          = {10.1109/IDSTA66210.2025.11202850},
}

@article{Alami2025CodeReview,
  title        = {Human and Machine: How Software Engineers Perceive and Engage with AI-Assisted Code Reviews Compared to Their Peers},
  author       = {Alami, Adam and Ernst, Neil A.},
  journal      = {arXiv preprint},
  year         = {2025},
  url          = {https://arxiv.org/abs/2501.02092},
}

@article{khattab2023dspy,
  title={Dspy: Compiling declarative language model calls into self-improving pipelines},
  author={Khattab, Omar and Singhvi, Arnav and Maheshwari, Paridhi and Zhang, Zhiyuan and Santhanam, Keshav and Vardhamanan, Sri and Haq, Saiful and Sharma, Ashutosh and Joshi, Thomas T and Moazam, Hanna and others},
  journal={arXiv preprint arXiv:2310.03714},
  year={2023}
}

@inproceedings{tao2024harnessing,
  title={Harnessing llms for api interactions: A framework for classification and synthetic data generation},
  author={Tao, Chunliang and Fan, Xiaojing and Yang, Yahe},
  booktitle={2024 5th International Conference on Computers and Artificial Intelligence Technology (CAIT)},
  pages={628--634},
  year={2024},
  organization={IEEE}
}

@article{bandlamudi2025framework,
  title={A framework for testing and adapting rest apis as llm tools},
  author={Bandlamudi, Jayachandu and Chaudhuri, Ritwik and Gantayat, Neelamadhav and Mukherjee, Kushal and Agarwal, Prerna and Sindhgatta, Renuka and Mehta, Sameep},
  journal={arXiv preprint arXiv:2504.15546},
  year={2025}
}

@inproceedings{kim2024leveraging,
  title={Leveraging large language models to improve rest api testing},
  author={Kim, Myeongsoo and Stennett, Tyler and Shah, Dhruv and Sinha, Saurabh and Orso, Alessandro},
  booktitle={Proceedings of the 2024 ACM/IEEE 44th International Conference on Software Engineering: New Ideas and Emerging Results},
  pages={37--41},
  year={2024}
}

@article{chauhan2025llm,
  title={Llm-generated microservice implementations from restful api definitions},
  author={Chauhan, Saurabh and Rasheed, Zeeshan and Sami, Abdul Malik and Zhang, Zheying and Rasku, Jussi and Kemell, Kai-Kristian and Abrahamsson, Pekka},
  journal={arXiv:2502.09766},
  year={2025}
}

@article{deng2025lrasgen,
  title={LRASGen: LLM-based RESTful API Specification Generation},
  author={Deng, Sida and Huang, Rubing and Zhang, Man and Cui, Chenhui and Towey, Dave and Wang, Rongcun},
  journal={arXiv preprint arXiv:2504.16833},
  year={2025}
}

@article{
doi:10.1073/pnas.1807184115,
author = {Jeffrey Heer },
title = {Agency plus automation: Designing artificial intelligence into interactive systems},
journal = {Proceedings of the National Academy of Sciences},
volume = {116},
number = {6},
pages = {1844-1850},
year = {2019},
}

@article{dellermann2021future,
  title={The future of human-AI collaboration: a taxonomy of design knowledge for hybrid intelligence systems},
  author={Dellermann, Dominik and Calma, Adrian and Lipusch, Nikolaus and Weber, Thorsten and Weigel, Sascha and Ebel, Philipp},
  journal={arXiv preprint arXiv:2105.03354},
  year={2021}
}

@inproceedings{amershi2019guidelines,
  title={Guidelines for human-AI interaction},
  author={Amershi, Saleema and Weld, Dan and Vorvoreanu, Mihaela and Fourney, Adam and Nushi, Besmira and Collisson, Penny and Suh, Jina and Iqbal, Shamsi and Bennett, Paul N and Inkpen, Kori and others},
  booktitle={Proceedings of the 2019 chi conference on human factors in computing systems},
  pages={1--13},
  year={2019}
}

@inproceedings{subramonyam2025prototyping,
  title={Prototyping with prompts: Emerging approaches and challenges in generative ai design for collaborative software teams},
  author={Subramonyam, Hari and Thakkar, Divy and Ku, Andrew and Dieber, Juergen and Sinha, Anoop K},
  booktitle={Proceedings of the 2025 CHI Conference on Human Factors in Computing Systems},
  pages={1--22},
  year={2025}
}

@article{piorkowski2021ai,
  title={How ai developers overcome communication challenges in a multidisciplinary team: A case study},
  author={Piorkowski, David and Park, Soya and Wang, April Yi and Wang, Dakuo and Muller, Michael and Portnoy, Felix},
  journal={Proceedings of the ACM on human-computer interaction},
  volume={5},
  number={CSCW1},
  pages={1--25},
  year={2021},
  publisher={ACM New York, NY, USA}
}

@inproceedings{Busquim2024Collaboration,
  author    = {Busquim, Gabriel and Ara{\'u}jo, Allysson Allex and Lima, Maria Julia and Kalinowski, Marcos},
  title     = {Towards Effective Collaboration between Software Engineers and Data Scientists developing Machine Learning-Enabled Systems},
  booktitle = {Brazilian Symposium on Software Engineering (SBES)},
  year      = {2024},
  address   = {Curitiba, Brazil},
  month     = {October}
}

@inproceedings{Kim2016DataScientists,
  author    = {Kim, Miryung and Zimmermann, Thomas and DeLine, Robert and Begel, Andrew},
  title     = {Characterizing the Roles of Data Scientists in a Large Software Company},
  booktitle = {ICSE'16: Proceedings of the 38th International Conference on Software Engineering},
  year      = {2016},
  address   = {Austin, TX, USA},
  month     = {May}
}

@inproceedings{li2017cross,
  title={Cross-disciplinary perspectives on collaborations with software engineers},
  author={Li, Paul Luo and Ko, Amy J and Begel, Andrew},
  booktitle={2017 IEEE/ACM 10th International Workshop on Cooperative and Human Aspects of Software Engineering (CHASE)},
  pages={2--8},
  year={2017},
  organization={IEEE}
}

@article{feng2025domains,
  title={When Domains Collide: An Activity Theory Exploration of Cross-Disciplinary Collaboration},
  author={Feng, Zixuan and Zimmermann, Thomas and Pisani, Lorenzo and Gooley, Christopher and Wander, Jeremiah and Sarma, Anita},
  journal={arXiv preprint arXiv:2506.20063},
  year={2025}
}

@InProceedings{10.1007/978-3-030-19034-7_14,
author="Lwakatare, Lucy Ellen
and Raj, Aiswarya
and Bosch, Jan
and Olsson, Helena Holmstr{\"o}m
and Crnkovic, Ivica",
editor="Kruchten, Philippe
and Fraser, Steven
and Coallier, Fran{\c{c}}ois",
title="A Taxonomy of Software Engineering Challenges for Machine Learning Systems: An Empirical Investigation",
booktitle="Agile Processes in Software Engineering and Extreme Programming",
year="2019",
pages="227--243",
isbn="978-3-030-19034-7"
}

@article{doi:10.1142/S0218194019500475,
author = {Pernst\r{a}l, Joakim and Feldt, R. and Gorschek, T. and Flor\'{e}n, D.},
title = {Communication Problems in Software Development — A Model and Its Industrial Application},
journal = {International Journal of Software Engineering and Knowledge Engineering},
volume = {29},
number = {10},
pages = {1497-1538},
year = {2019}
}

@article{heyn2023automotive,
  title={Automotive perception software development: An empirical investigation into data, annotation, and ecosystem challenges},
  author={Heyn, Hans-Martin and Habibullah, Khan Mohammad and Knauss, Eric and Horkoff, Jennifer and Borg, Markus and Knauss, Alessia and Li, Polly Jing},
  journal={arXiv preprint arXiv:2303.05947},
  year={2023}
}

@INPROCEEDINGS {6602469,
author = { Weber-Jahnke, Jens H. and Price, Morgan and Williams, James },
booktitle = { 2013 5th International Workshop on Software Engineering in Health Care (SEHC) },
title = {{ Software engineering in health care: Is it really different? And how to gain impact }},
year = {2013},
volume = {},
ISSN = {},
pages = {1-4},
doi = {10.1109/SEHC.2013.6602469},
url = {https://doi.ieeecomputersociety.org/10.1109/SEHC.2013.6602469},
publisher = {IEEE Computer Society},
address = {Los Alamitos, CA, USA},
month =May}

@article{sculley2015hidden,
  title={Hidden technical debt in machine learning systems},
  author={Sculley, David and Holt, Gary and Golovin, Daniel and Davydov, Eugene and Phillips, Todd and Ebner, Dietmar and Chaudhary, Vinay and Young, Michael and Crespo, Jean-Francois and Dennison, Dan},
  journal={Advances in neural information processing systems},
  volume={28},
  year={2015}
}

@book{10.5555/583119,
author = {Bennett, Stuart},
title = {A  History of Control Engineering 1930-1955},
year = {1993},
isbn = {0863412998},
publisher = {Peter Peregrinus},
address = {GBR},
edition = {1st}
}

@article{DBLP:journals/computer/Brooks87,
  author       = {Frederick P. Brooks Jr.},
  title        = {No Silver Bullet - Essence and Accidents of Software Engineering},
  journal      = {Computer},
  volume       = {20},
  number       = {4},
  pages        = {10--19},
  year         = {1987},
  url          = {https://doi.org/10.1109/MC.1987.1663532},
  doi          = {10.1109/MC.1987.1663532},
  bibsource    = {dblp computer science bibliography, https://dblp.org}
}

@inproceedings{10.1145/3377816.3381720,
author = {Casalnuovo, Casey and Barr, Earl T. and Dash, Santanu Kumar and Devanbu, Prem and Morgan, Emily},
title = {A theory of dual channel constraints},
year = {2020},
isbn = {9781450371261},
publisher = {Association for Computing Machinery},
address = {New York, NY, USA},
pages = {25–28},
numpages = {4},
location = {Seoul, South Korea},
series = {ICSE-NIER '20}
}

@misc{brown2020languagemodelsfewshotlearners,
      title={Language Models are Few-Shot Learners}, 
      author={Tom B. Brown and Benjamin Mann and Nick Ryder and Melanie Subbiah and Jared Kaplan and Prafulla Dhariwal and Arvind Neelakantan and Pranav Shyam and Girish Sastry and Amanda Askell and Sandhini Agarwal and Ariel Herbert-Voss and Gretchen Krueger and Tom Henighan and Rewon Child and Aditya Ramesh and Daniel M. Ziegler and Jeffrey Wu and Clemens Winter and Christopher Hesse and Mark Chen and Eric Sigler and Mateusz Litwin and Scott Gray and Benjamin Chess and Jack Clark and Christopher Berner and Sam McCandlish and Alec Radford and Ilya Sutskever and Dario Amodei},
      year={2020},
      eprint={2005.14165},
      archivePrefix={arXiv},
      primaryClass={cs.CL},
      url={https://arxiv.org/abs/2005.14165}, 
}

@misc{bubeck2023sparksartificialgeneralintelligence,
      title={Sparks of Artificial General Intelligence: Early experiments with GPT-4}, 
      author={S\'ebastien Bubeck and Varun Chandrasekaran and Ronen Eldan and Johannes Gehrke and Eric Horvitz and Ece Kamar and Peter Lee and Yin Tat Lee and Yuanzhi Li and Scott Lundberg and Harsha Nori and Hamid Palangi and Marco Tulio Ribeiro and Yi Zhang},
      year={2023},
      eprint={2303.12712}, 
}

@article{jaimovitch2023can,
  title={Can language models automate data wrangling?},
  author={Jaimovitch-L{\'o}pez, Gonzalo and Ferri, C{\`e}sar and Hern{\'a}ndez-Orallo, Jos{\'e} and Mart{\'\i}nez-Plumed, Fernando and Ram{\'\i}rez-Quintana, Mar{\'\i}a Jos{\'e}},
  journal={Machine Learning},
  volume={112},
  number={6},
  pages={2053--2082},
  year={2023},
  publisher={Springer}
}

@article{narayan2022can,
  title={Can foundation models wrangle your data?},
  author={Narayan, Avanika and Chami, Ines and Orr, Laurel and Arora, Simran and R{\'e}, Christopher},
  journal={arXiv:2205.09911},
  year={2022}
}

@inproceedings{DBLP:conf/icml/Beurer-Kellner024,
  author       = {Luca Beurer{-}Kellner and
                  Marc Fischer and
                  Martin T. Vechev},
  title        = {Guiding LLMs The Right Way: Fast, Non-Invasive Constrained Generation},
  booktitle    = {Forty-first International Conference on Machine Learning, {ICML} 2024,
                  Vienna, Austria, July 21-27, 2024},
  publisher    = {OpenReview.net},
  year         = {2024},
  url          = {https://openreview.net/forum?id=pXaEYzrFae},
  bibsource    = {dblp computer science bibliography, https://dblp.org}
}

@misc{ugare2024syncodellmgenerationgrammar,
      title={SynCode: LLM Generation with Grammar Augmentation}, 
      author={Shubham Ugare and Tarun Suresh and Hangoo Kang and Sasa Misailovic and Gagandeep Singh},
      year={2024},
      eprint={2403.01632},
      archivePrefix={arXiv},
      primaryClass={cs.LG},
      url={https://arxiv.org/abs/2403.01632}, 
}

@misc{park2024grammaraligneddecoding,
      title={Grammar-Aligned Decoding}, 
      author={Kanghee Park and Jiayu Wang and Taylor Berg-Kirkpatrick and Nadia Polikarpova and Loris D'Antoni},
      year={2024},
      eprint={2405.21047},
      archivePrefix={arXiv},
      primaryClass={cs.AI}, 
}

@misc{wang2025automatingcompletesoftwaretest,
      title={Automating a Complete Software Test Process Using LLMs: An Automotive Case Study}, 
      author={Shuai Wang and Yinan Yu and Robert Feldt and Dhasarathy Parthasarathy},
      year={2025},
      eprint={2502.04008},
      archivePrefix={arXiv},
      primaryClass={cs.SE},
      url={https://arxiv.org/abs/2502.04008}, 
}

@article{KBE1,
author = {Chen Zheng and Matthieu Bricogne and Julien Le Duigou and Peter Hehenberger and Benoit Eynard},
title ={Knowledge-based engineering for multidisciplinary systems: Integrated design based on interface model},
journal = {Concurrent Engineering},
year = {2018}
}

@article{KBE2,
title = {Management of Heterogeneous Information for Integrated Design of Multidisciplinary Systems},
journal = {Procedia CIRP},
volume = {60},
pages = {320-325},
year = {2017},
note = {Complex Systems Engineering and Development Proceedings of the 27th CIRP Design Conference Cranfield University, UK 10th – 12th May 2017},
issn = {2212-8271},
doi = {https://doi.org/10.1016/j.procir.2017.02.020},
url = {https://www.sciencedirect.com/science/article/pii/S2212827117301166},
author = {Benjamin Guerineau and Chen Zheng and Matthieu Bricogne and Alexandre Durupt and Louis Rivest and Harvey Rowson and Benoît Eynard}
}
\end{document}